\documentclass[epj]{webofc}
\usepackage[varg]{txfonts}   
\woctitle{International Conference on New Frontiers in Physics 2013}
 \newcommand\la{\langle}
 \newcommand\ra{\rangle}
 \newcommand\beq{\begin{equation}}
 
 \newcommand\eeq{\end{equation}}
 \newcommand\beqn{\begin{eqnarray}}
 \newcommand\eeqn{\end{eqnarray}}
 \newcommand\GeV{{\rm GeV}}
 
\def\halftext{.47\textwidth}

\def\Im{\,\mbox{Im}\,}

\def\mb{\,\mbox{mb}}
\def\fm{\,\mbox{fm}}
\def\GeV{\,\mbox{GeV}}
\def\TeV{\,\mbox{TeV}}

\def\lsim{\mathrel{\rlap{\lower4pt\hbox{\hskip1pt$\sim$}}
    \raise1pt\hbox{$<$}}}         
\def\gsim{\mathrel{\rlap{\lower4pt\hbox{\hskip1pt$\sim$}}
    \raise1pt\hbox{$>$}}}         

\def\la{\langle}
\def\ra{\rangle}

\begin{document}
\title{\boldmath Quenching of high-$p_T$ hadrons: a non-energy-loss scenario}
%
%

\author{B. Z. Kopeliovich\inst{1}\fnsep\thanks{\email{boris.kopeliovich@usm.cl}} \and
        J. Nemchik\inst{2,3}
              \and
        I. K. Potashnikova\inst{1}
           \and
             Iv\'an Schmidt\inst{1}
}

\institute{Departamento de F\'{\i}sica,
Universidad T\'ecnica Federico Santa Mar\'{\i}a;\\
Centro Cient\'ifico-Tecnol\'ogico de Valpara\'{\i}so,
Avda. Espa\~na 1680, Valpara\'{\i}so, Chile
\and
           Czech Technical University in Prague,
FNSPE, B\v rehov\'a 7,
11519 Prague, Czech Republic
\and
           Institute of Experimental Physics SAS, Watsonova 47,
04001 Ko\v sice, Slovakia
          }

\abstract{A parton produced with a high transverse momentum in a hard collision is regenerating its color field, intensively radiating gluons and losing energy. This process cannot last long, if it ends up with production of a leading hadron carrying the main fraction $z_h$ of the initial parton momentum. So energy conservation imposes severe constraints on the length scale of production of a single hadron with high $p_T$. As a result, the main reason for hadron quenching observed in heavy ion collisions, is not energy loss, but attenuation of the produced colorless dipole in the created dense medium. The latter mechanism, calculated with the path-integral method, explains well the observed suppression of light hadrons and the elliptic flow in a wide range of energies, from the lowest energy of RHIC up to LHC, and in a wide range of transverse momenta. The values of the transport coefficient extracted from data range within $1$-$2\GeV^2\!/\!\fm$, dependent on energy, and agree well with the theoretical expectations.
}
\maketitle

\section{Intoduction}

Single hadrons produced with high transverse momenta  in heavy ion collisions
at high energies of RHIC \cite{phenix,star} and LHC \cite{alice-new,cms-new1,cms-new2} turn out to be appreciably suppressed compared with $pp$ collisions. 
While there is a consensus about the source of the suppression, which is final-state interaction with the co-moving medium, created in the collision,
the mechanisms of the interaction is still under debate.

The popular interpretation of the observed  suppression of high-$p_T$ hadrons is the loss of energy by a parton propagating through the medium created in the collision. The perturbative radiative energy loss
is caused by the "wiggling" of the parton trajectory due to multiple interactions in the medium. Every time, when the parton gets a kick from a scattering in the medium, a new portion of its color field is shaken off. 
The loss of energy induced by multiple interactions is naturally related to the broadening of the parton transverse  (relative to its trajectory, i.e. to $\vec p_T$) momentum $k_T$ \cite{bdmps},
\beq
\frac{dE}{dL}=-\frac{3\alpha_s}{4}\,\Delta k_T^2(L)=
-\frac{3\alpha_s}{4}\int\limits_0^L dl\, \hat q(l),
\label{100}
\eeq
where $\hat q(l)$ is the rate of broadening $\Delta k_T^2$, which may vary with $l$ along the parton trajectory,
\beq
\hat q(l)=\frac{d\Delta k_T^2}{dl}.
\label{120}
\eeq

Although it is natural to expect that dissipation of energy by the parton in a medium should suppress production of leading hadrons, realization of this idea in detail raises many questions.
In particular, one usually assumes that energy loss results in a shift in the argument of the fragmentation function, $z_h\Rightarrow z_H+\Delta z_h$. This could be true, if hadronization of the parton started outside the medium. However it starts right away after the hard collision, and the main part of gluon radiation occurs on a short distance (see section~\ref{eloss}).

Another assumption, which the energy loss scenario relies upon and which has never been  justified, is that the path length of the parton in the medium is always longer than the span of the medium.
So the colorless hadronic state, which does not radiate energy any more and is eventually detected, is produced outside the medium. The validity of this assumption should be investigated and the path length available for hadronization should be evaluated.
This problem
was debated in \cite{within} on a more certain situation of semi-inclusive deep-inelastic scattering,
as well as within dynamical models of hadronization \cite{pert-hadr, jet-lag} providing solid constraints on the above assumption.

\begin{figure}[htb]
\parbox{\halftext}{
\centerline{\includegraphics[width=6 cm]{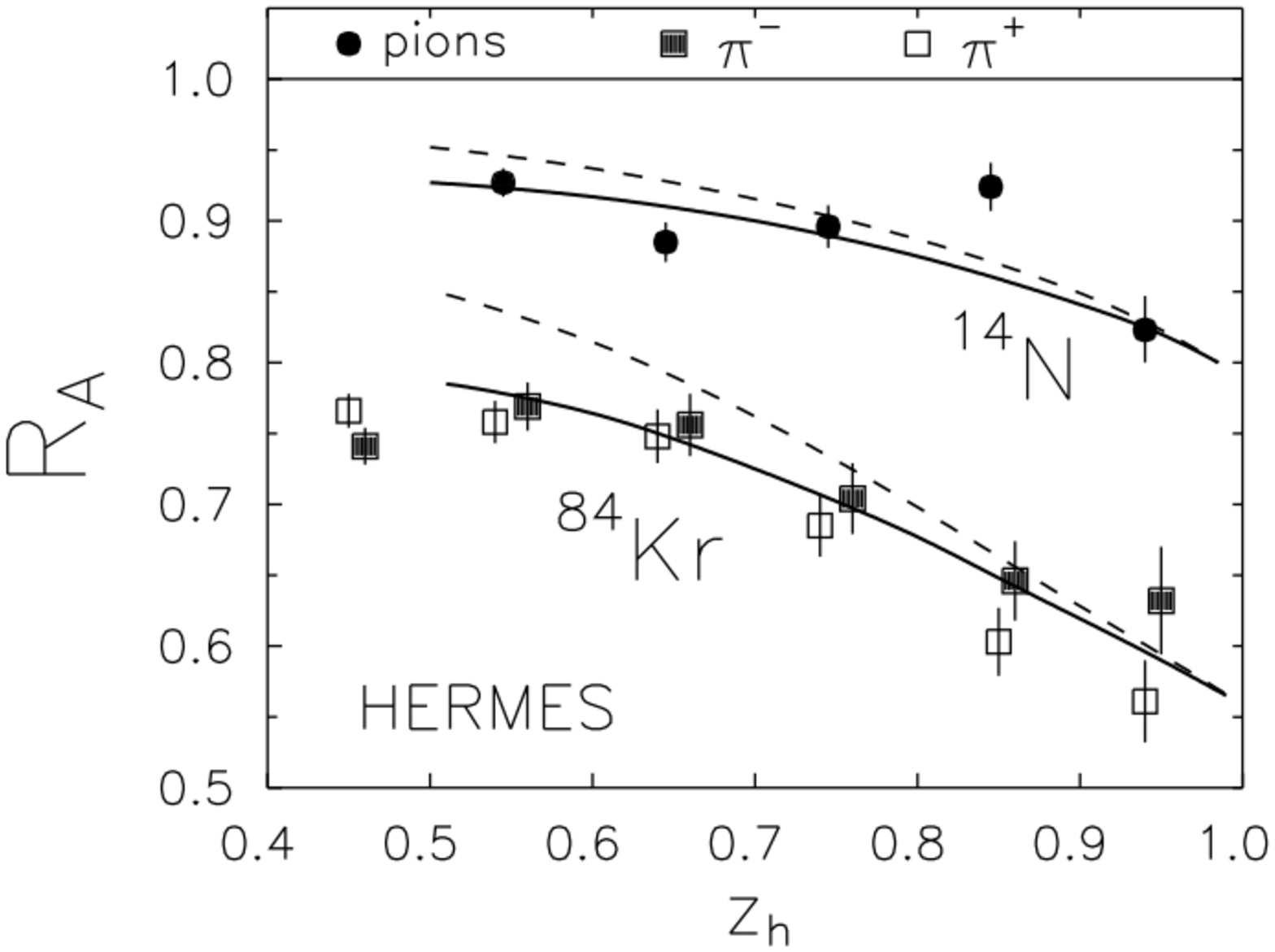}}
\caption{Comparison of the predicted \cite{knp,within} $z_h$-dependence of the nuclear suppression factor in inclusive electroproduction of pions with HERMES data \cite{hermes1,hermes2}.
Solid and dashed curves show the results with included or neglected energy loss corrections, respectively.}
\label{fig:mine}}
\hfill
\parbox{\halftext}{\vspace*{-3mm}
\centerline{\includegraphics[width=5 cm]{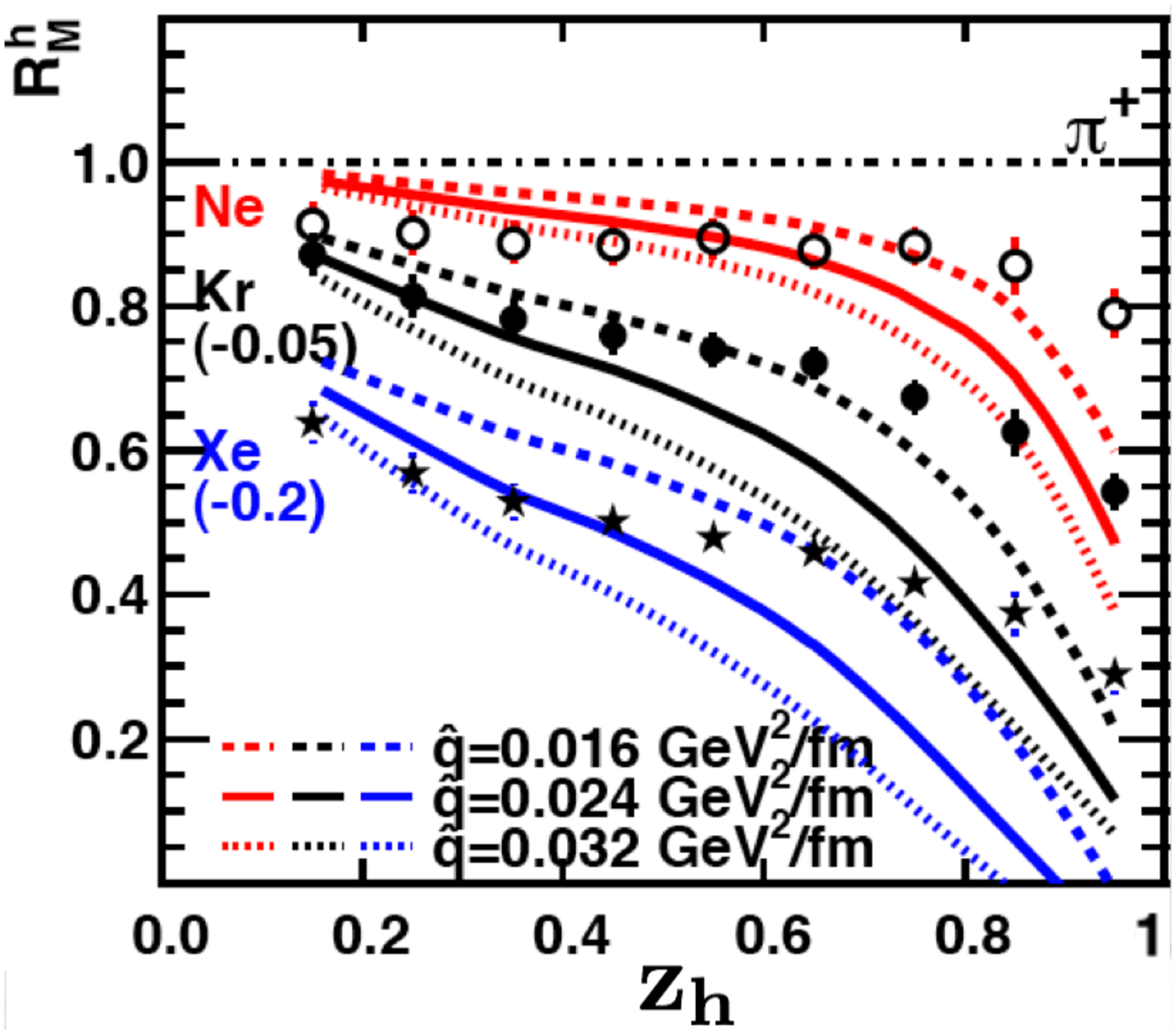}}
\caption{The results of description of the HERMES results by the model \cite{wang-sidis} based on the energy loss scenario, i.e. assuming a long time of hadronization. Several values of the transport coefficient are tested. }
\label{fig:wang}}
\end{figure}

Besides, theoretical arguments, some experimental data is also difficult to explain within the energy loss scenario. In particular, production of hadrons in semi-inclusive deeply inelastic scattering (SIDIS) offer a rigorous test for in-medium hadronization model in much more certain environment than in heavy ion collisions. Indeed in SIDIS off nuclei the medium density and geometry are well known and time independent; the fractional momentum $z_h$ (the argument of the fragmentation function) of the detected hadron is measured.
Measurements performed in the HERMES experiment \cite{hermes1,hermes2}
well confirmed the predictions \cite{knp} made five years prior the measurements,
within a model, which include evaluation of the hadronization length, which was found rather short. The comparison with data shown in Fig.~\ref{fig:mine} \cite{within} demonstrate a good agreement.
 \vspace*{-5mm}
 
 On the other hand the attempts to explain the same results of HERMES within the energy loss scenario were not successful. An example of comparison the model \cite{wang-sidis} with data \cite{hermes2} depicted in Fig.~\ref{fig:wang} show that the model fails to explain the data at large $z_h>0.5$, which dominates high-$p_T$ hadron production in heavy ion collision
 (see Fig.~\ref{fig:mean-zh}). Even adjustment of the transport coefficient $\hat q$ (actually well known for the cold nuclear matter \cite{broad,domdey}) did not help.
 
\begin{figure}[htb]
\parbox{\halftext}{
\centerline{\includegraphics[width=6.5 cm]{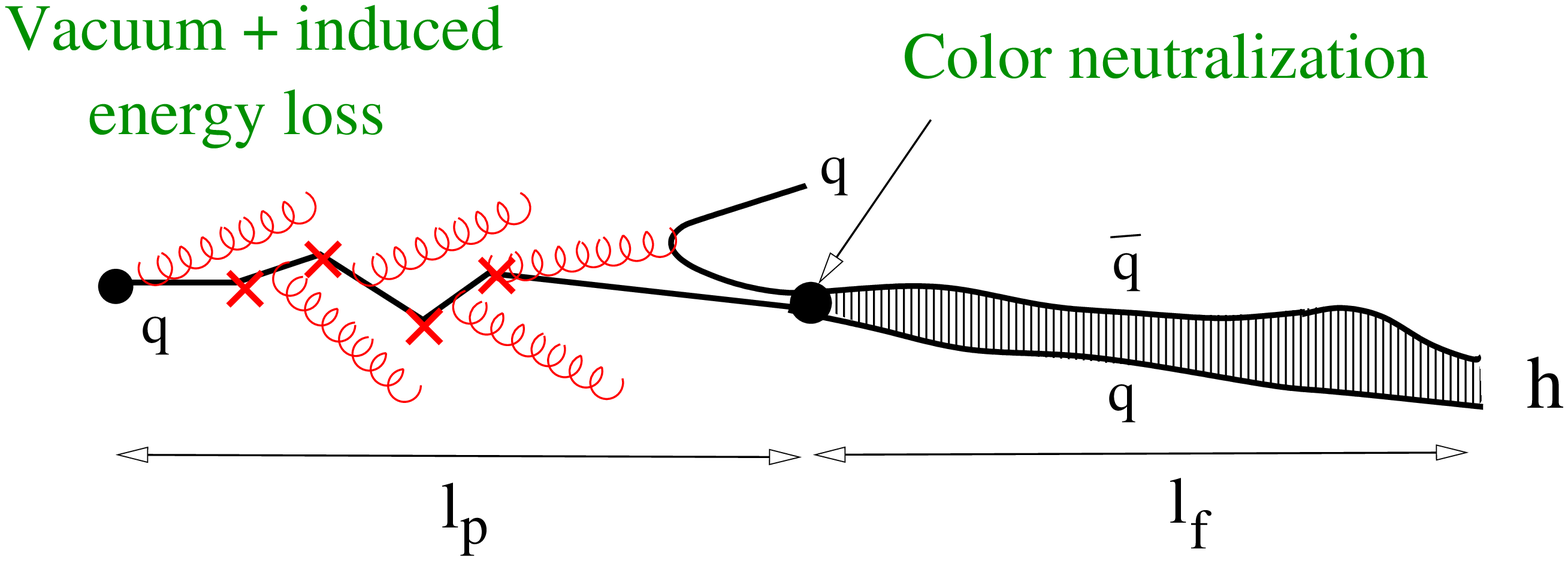}}
\caption{Space-time development of hadronization of a highly virtual quark producing a leading hadron
carrying a substantial fraction $z_h$ of the initial  light-cone momentum.}
\label{fig:space-time}}
\hfill
\parbox{\halftext}{
\centerline{\includegraphics[width=5 cm]{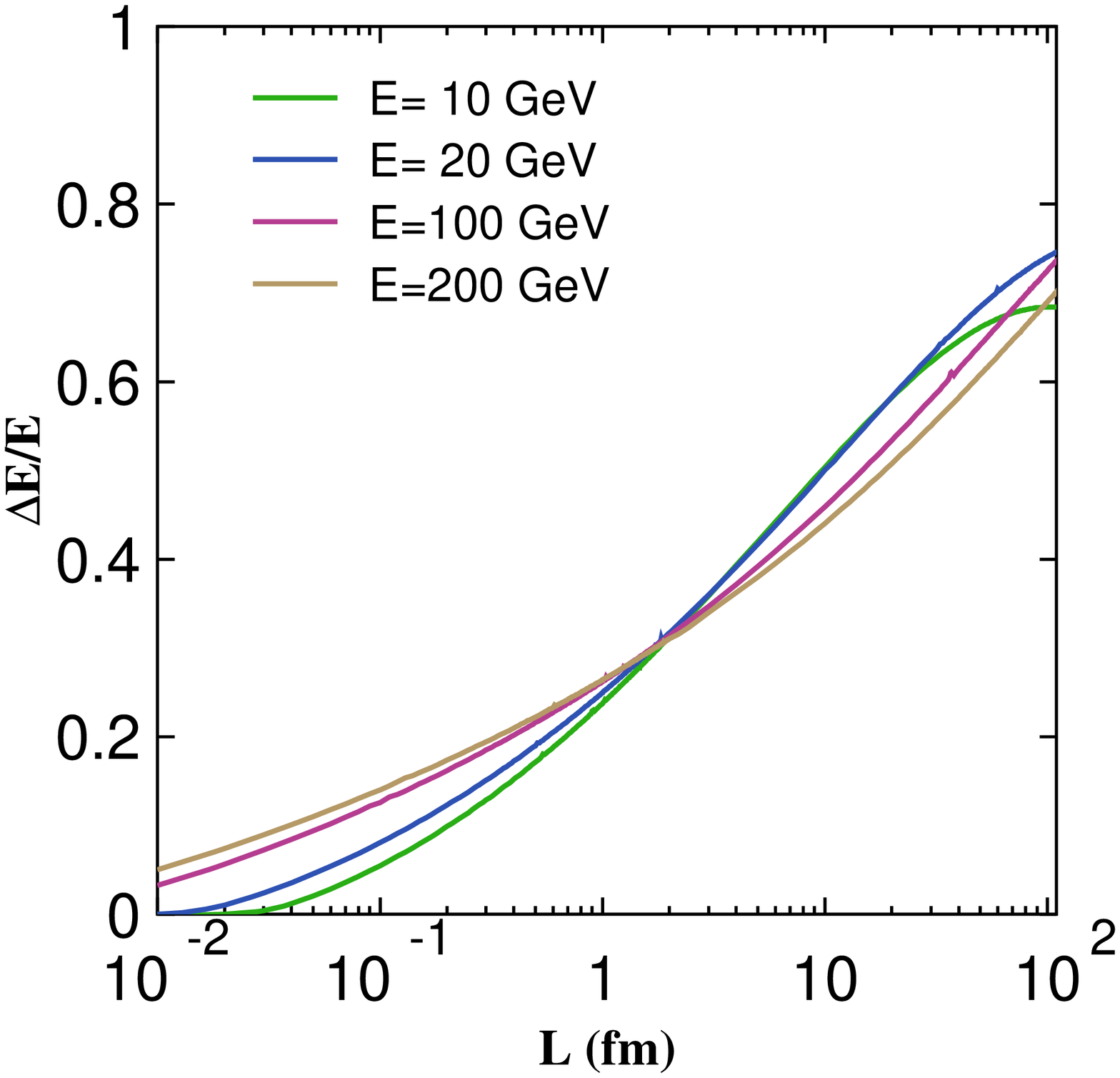}}
\caption{ Fractional  energy loss by a quark with different initial energies in vacuum vs path length $L$. }\label{fig:e-loss}}
\end{figure}

Thus, the assumption of a long hadronization length should be checked thoroughly. 
The main space-time scales of the in-medium hadronization process are indicated schematically in Fig.~\ref{fig:space-time}.
We call production length $l_p$  the distance at which color neutralization occurs and a colorless "pre-hadron" (a colorless state, which does not have any certain mass) is produced and starts developing a wave function. The latter process is characterized by the formation length, which usually is rather long, $l_f\sim 2E/(m_{h^*}^2-m_h^2)$ \cite{kz91,within}.
In what follows we concentrate on the production length scale $l_p$, which is the only path available for energy loss. 

Notice that the question whether the hadronization ends up within or without the medium, might have no certain answer. This is a typical quantum-mechanical uncertainty:
the production amplitudes with different values of $l_p$ interfere.
Such interference was evaluated in \cite{avt} for an example of a SIDIS process and found
to be a considerable effect. However, an extension of those results to a hot medium is a challenge, so what follows we neglect the interference and employ the standard semi-classical space-time pattern of the production process. 

\section{Radiative energy loss in vacuum}\label{eloss}

First of all, one should discriminate between vacuum and medium-induced radiative energy loss. High-$p_T$ partons radiate gluons and dissipate energy even in vacuum, and the corresponding rate of energy loss may considerably exceed the medium-induced 
value Eq.~(\ref{100}), because the former is caused by a hard collision.

\subsection{\boldmath Regeneration of the color field of a high-$p_T$ parton}

High-$p_T$ scattering of partons leads to an intensive gluon
radiation in forward-backward directions, which is related to the
initial color field of the partons, shaken off due to the strong
acceleration caused by the hard  collision. The
Weitz\"acker-Williams gluons accompanying the colliding partons do not survive
the hard interaction and lose coherence up to transverse frequencies
$k\lsim p_T$. Therefore, the produced high-$p_T$ parton is lacking
this part of the field and starts regenerating it via radiation
of a new cone of gluons, which are aligned along the new direction.
One can explicitly see the two cones of radiation in the Born approximation calculated in \cite{gunion-bertsch}.
This process lasts a long time proportional to the jet energy
($E\approx p_T$), since the coherence length (or time) $l_g$ of gluon radiation depends on the gluon fractional light-cone momentum $x$ and its transverse momentum $k$ relative to the jet axis as
(to be concrete we assume that the jet is initiated by a quark), 
\beq
l_g=\frac{2E}{M_{qg}^2-m_q^2}= \frac{2Ex(1-x)}{k^2+x^2\,m_q^2}.
\label{140} 
\eeq 
Here 
$M_{qg}$ is the invariant mass of
the recoil quark and radiated gluon.

\begin{figure}[htb]
\parbox{\halftext}{
\vspace{3mm}
\centerline{\includegraphics[width=5.5 cm]{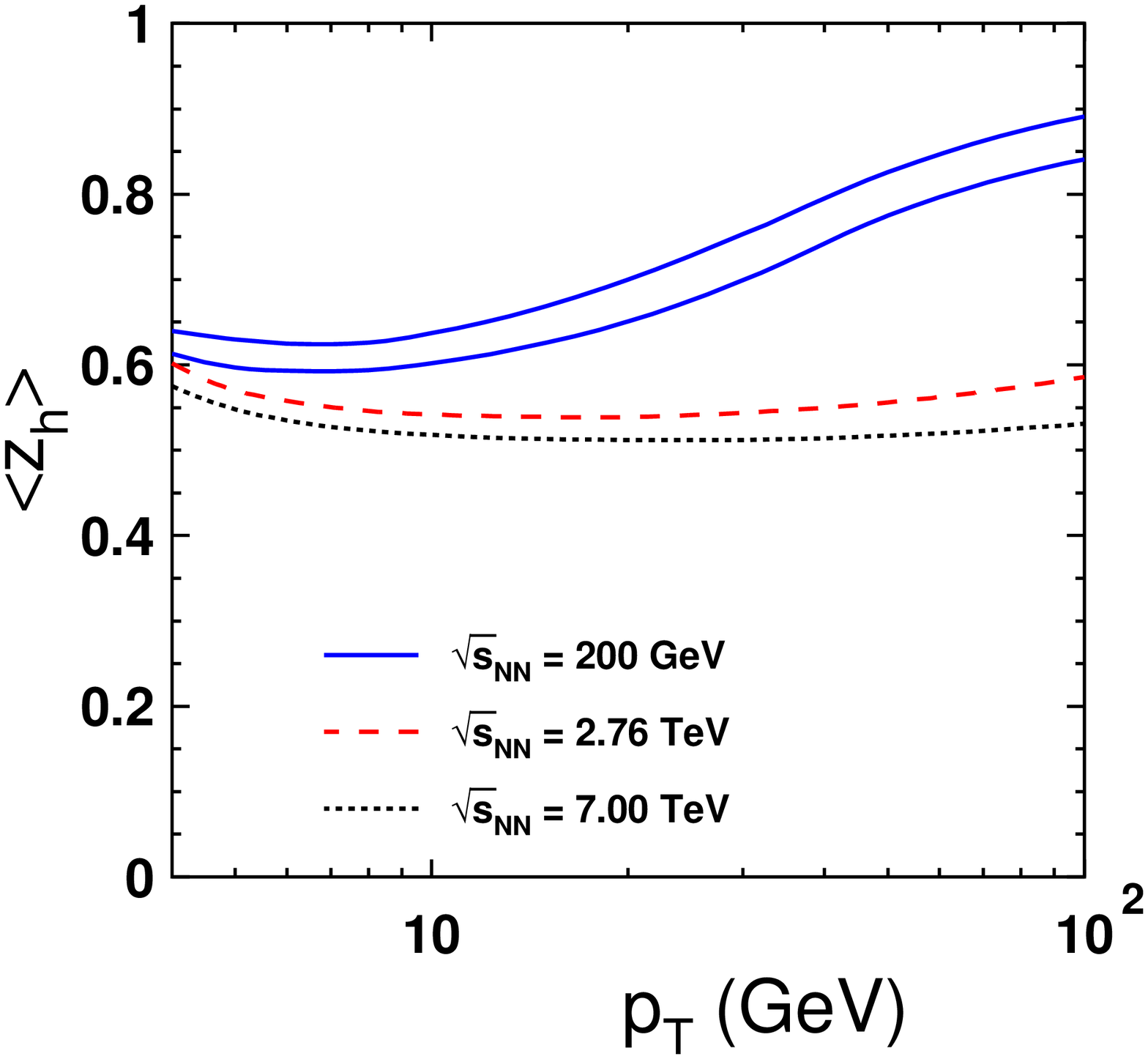}}
\caption{The mean fraction $\la z_h\ra$  of the jet energy carried by a hadron detected with transverse momentum $p_T$. The calculations are performed for collision energies $\sqrt{s}=0.2,\ 2.76$ and $7\TeV$.}
\label{fig:mean-zh}}
\hfill
\parbox{\halftext}{
\centerline{\includegraphics[width=10.7 cm]{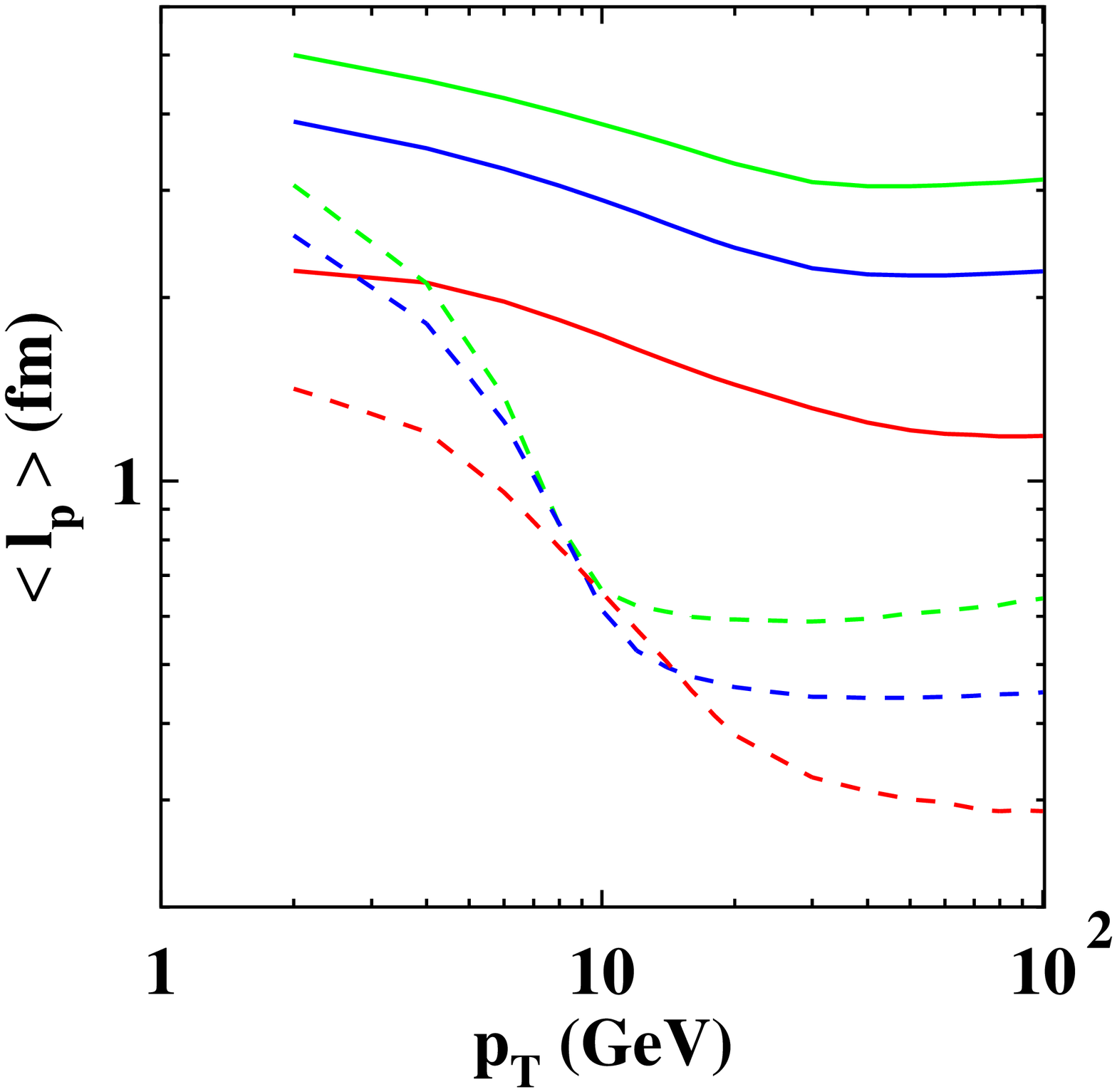}}
\caption{ The mean production length as function of energy for quark (solid curves) and gluon (dashed curves) jets.
In both cases the curves are calculated at $z_h=0.5,\ 0.7,\ 0.9$ (from top to bottom). } \label{fig:mean-lp}}
\end{figure}

One can trace how much energy is lost over the path length $L$ via gluons which have lost coherence,
i.e. were radiated, during this time interval,
 \beq
\frac{\Delta E(L)}{E} =
\int\limits_{\Lambda^2}^{Q^2}
dk^2\int\limits_0^1 dx\,x\,
\frac{dn_g}{dx\,dk^2}
\Theta(L-l_g),
\label{160}
 \eeq
 where $Q\sim p_T$ is the initial quark virtuality; the infra-red cutoff is fixed at $\Lambda=0.2\GeV$.
 The radiation spectrum reads
 \beq
\frac{dn_g}{dx\,dk^2} =
\frac{2\alpha_s(k^2)}{3\pi\,x}\,
\frac{k^2[1+(1-x)^2]}{[k^2+x^2m_q^2]^2}
\label{180}
 \eeq
 A few examples of fractional vacuum energy loss by a quark vs distance from the hard collision is depicted in Fig.~\ref{fig:e-loss} for  initial energies $10,\  20,\ 100,\ 200\GeV$ (compare with heavy flavors in \cite{eloss}).
The rate of energy dissipation is considerable and energy conservation may become an issue for a long path length, if one wants to produce a leading hadron. Indeed, the production rate of high-$p_T$ hadrons
comes from a convolution of the the parton distributions in the colliding hadrons (which suppresses large fractional momenta $x$, i.e. high $p_T$), with the transverse momentum distribution in the hard parton collisions (also suppresses large $p_T$), and with the fragmentation function $D(z_h)$ of the produced parton.
The latter has maximum at small $z_h\ll1$, which, however, is strongly suppressed by the convolution, pushing the maximum towards large values of $z_h$. Numerical results of the convolution for the mean value $\la z_h\ra$ \cite{my-alice,ct-eloss} are depicted in Fig.~\ref{fig:mean-zh}, separately for quark and gluon jets (upper and bottom solid curves) and at different energies, $\sqrt{s}=200,\ 2760$ and $7000\GeV$.
We see that the lower the collision energy is, the larger is $\la z_h\ra$, especially at high $p_T$, because the parton $k_T$ distribution gets steeper. In the energy range of the LHC the magnitude of $\la z_h\ra$ practically saturates as function of $\sqrt{s}$ and $p_T$, and becomes indistinguishable for quark and gluonic jets.

It is worth emphasizing the difference between inclusive production of a high-$p_T$ hadron
and a high-$p_T$ jet. In the former case, according to the above consideration, the detected hadron carries the main fractional light-cone momentum $z_h$ of the jet, which it originates from.
In the latter case, if only the whole jet transverse momentum $p_T^{jet}$ is required to be large, and no other constraints are imposed, the fractional momenta of hadrons in the jet are typically very low, so energy conservation energy conservation does not imply any severe constraints on the hadronization time scale, which rises with $p_T$ and may be long.

\subsection{How long does it take to produce a hadron?}

Apparently, production of a hadron with fractional momentum $z_h$ becomes impossible after the parton initiating the jet, radiates a substantial fraction of its initial energy, $\Delta E/E>1-z_h$. Thus, energy conservation imposes an upper bound for the production length $l_p$.
Figures \ref{fig:e-loss} and \ref{fig:mean-zh} show that such a maximal value of $l_p$ is rather short. Remarkably, its value is nearly independent of $p_T$. This might seem to be in contradiction with the Lorentz factor, which should lead to $l_p\propto p_T$. However, the intensity of gluon radiation and the rate of energy loss also rise, approximately as $p_T^2$,
what leads to an opposite effect of $l_p$ reduction.
 
More precisely the $p_T$ dependence of $\la l_p\ra$ can be derived  within a dynamic model for hadronization. It was done in Ref.~\cite{jet-lag,my-alice} employing the model of perturbative hadronization \cite{pert-hadr}.
Some numerical results are plotted in Fig.~\ref{fig:mean-lp} for fragmentation of quarks and gluons by solid and dashed curves respectively. We see that the mean production length is rather short and slowly decreases with 
$p_T$. The production length for gluon jets is shorter, because of a more intensive vacuum energy loss and a stronger Sudakov suppression, which leads to a reduction of $\la l_p\ra$. 
 
The  production length in Fig.~\ref{fig:mean-lp}  demonstrates a trend to decrease with $p_T$, which is in variance with the naive expectation of a rise due to the Lorentz factor. 
As was explained above, this happens due to the growing virtuality and radiative dissipation of energy.

Notice that we took into consideration so far only the dissipation of energy in vacuum. Apparently, adding the medium induced 
energy loss one can only enhance the energy deficit and make the production length even shorter.

\section{Propagation of color dipoles in a dense medium}

As was discussed above, the intensive gluon radiation following the hard parton interaction, must stop in a short while (see Fig.~\ref{fig:mean-lp}) in order to be able to fragment into a hadron with large $z_h$ (see Fig.~\ref{fig:mean-zh}) and to obey the energy conservation constraints. The gluon radiation can stop only by means of color neutralization and production of a colorless dipole. In vacuum (e.g. in $pp$ collision) the matrix element for a hadron production contains a direct projection of the initial distribution amplitude of the dipole separation to the hadron wave function,
\beq 
A_{{\rm vac}} \propto 
\int\limits_0^1d\alpha\int d^2r\,
\Psi_h^\dagger(\vec r,\alpha)
\Psi_{in}(\vec r,\alpha),
\label{200}
\eeq
where $\Psi_h$ and $\Psi_{in}$ are the light-cone wave function of the hadron and the distribution amplitude of the produced dipole, respectively. We use the mixed representation of transverse dipole separation $\vec r$ and fractional light-cone momentum $\alpha$ carried by the quark.

In the case of production inside a medium, the produced dipole has to survive the propagation through the medium, i.e. to experience no inelastic collisions. The inelastic cross section depends on the dipole separation, which is fluctuating during the propagation. The effective way to solve this problem on a strict quantum-mechanical ground is the path integral method \cite{kz91,ct-eloss}.
In the case of in-medium production the matrix element Eq.~(\ref{200}) is modified to,
\beq
A_{{\rm med}} \propto 
\int\limits_0^1d\alpha\int d^2r_1 d^2r_2\,
\Psi_h^\dagger(\vec r_2,\alpha)\,
G_{\bar qq}(l_1,\vec r_1;l_2,\vec r_2)\,
\Psi_{in}(\vec r_1,\alpha).
\label{220}
\eeq
Here $G_{\bar qq}(l_1,\vec r_1;l_2,\vec r_2)$ is the Green function describing propagation of the dipole between longitudinal coordinates $l_1$ and $l_2$ with initial and final separations $\vec r_1$ and $\vec r_2$ respectively. It satisfies the two-dimensional Schr\"odinger equation
\cite{kst1,kst2,krt,zakharov},
\beq
i\frac{d}{dl_2}\,G_{\bar qq}(l_1,\vec r_1;l_2,\vec r_2) =
\left[
\frac{m_q^2 - \Delta_{r_2}}{2\,p_T\,\alpha\,(1-\alpha)}
-V_{\bar qq}(l_2,\vec r_2)\right]\,
G_{\bar qq}(l_1,\vec r_1;l_2,\vec r_2)
\label{240}
\eeq
The  first term in square brackets plays role of kinetic energy in Schr\"odinger equation, while the imaginary part of the light-cone potential $V_{\bar qq}(l_2,\vec r_2)$ is responsible for absorption in the medium. The relation  between the rate of broadening and the dipole cross section derived in \cite{jkt}, allows to present the imaginary part of the potential as \cite{psi,my-alice},
\beq
\Im V_{\bar qq}(l,\vec r) = -{1\over4}\,\hat q(l)\,r^2.
\label{260}
\eeq
Thus, color transparency \cite{zkl} controls dipole attenuation in a medium.

The real part of the potential describes the nonperturbative interaction between $q$ and $\bar q$ in the dipole \cite{kst2,VM}. However, it 
should not affect much the dipole evolution on the initial perturbative stage of development.
Therefore we will treat the $\bar qq$ as free noninteracting partons. 

Measurements of the suppression of high-$p_T$ hadrons in heavy ion collisions can provide precious information about the properties of the created hot matter. 
One can quantify those properties by the value of the transport coefficient $\hat q$, which is proportional to the medium density.
The  latter is time dependent, and
is assumed to dilute as  $\rho(t)=\rho_0\,t_0/t$ due to the
longitudinal expansion of the produced medium. 
We adopt the popular parametrization \cite{frankfurt} of  the transport coefficient, which depends on impact parameter and path length (time) as,
\beq 
\hat q(l,\vec b,\vec\tau)=\frac{\hat
q_0\,l_0}{l}\, \frac{n_{part}(\vec b,\vec\tau)}{n_{part}(0,0)},
\label{280} 
\eeq
where $n_{part}(\vec b,\vec\tau)$ is the number of participants;
$\hat q_0$ corresponds to the
maximum medium density produced at impact parameter $\tau=0$ in
central collisions ($b=0$) at the time $t=t_0=l_0$ after
the collision. In what follows we treat the transport coefficient
$\hat q_0$ as an adjusted parameter.

\section{Results at the energies of LHC}

Now we are in a position to calculate the suppression factor $R_{AB}(b,p_T)$
for high-$p_T$ hadrons produced in nuclear $A$-$B$ collision with impact
parameter $b$. The suppression orrurs due to the difference between the matrix elements (\ref{200}) and (\ref{220}), so we get,
\beq
R_{AB}(\vec{b},p_T) =
\frac{\int d^2\tau \,T_A(\tau)T_B(\vec b-\vec\tau)
\int\limits_0^{2\pi}\frac{d\phi}{2\pi} 
\left|
\int\limits_0^1d\alpha\int d^2r_1 d^2r_2\,
\Psi_h^\dagger(\vec r_2,\alpha)
G_{\bar qq}(l_1,\vec r_1;l_2,\vec r_2)
\Psi_{in}(\vec r_1,\alpha)\right|^2
}{T_{AB}(b)
\left|\int\limits_0^1d\alpha\int d^2r\,
\Psi_h^\dagger(\vec r,\alpha)
\Psi_{in}(\vec r,\alpha)\right|^2
}.
\label{300}
\eeq
where $T_{AB}=\int d^2\tau\,T_A(b)T_B(\vec b-\vec\tau)$;
$\phi$ is the azimuthal angle between the dipole trajectory and reaction plane (impact vector $\vec b$). 

We also included medium-induced radiative energy loss during the short path from $l=l_0\sim1\fm$ to $l=l_p$ (if $l_p>l_0$), where the parton experiences
multiple interactions, which induce extra radiation of gluons and additional loss of energy
\cite{bdmps},
\beq
\Delta E_{\rm med}=\frac{3\alpha_s}{4}\,
\Theta(l_p-l_0)
\int\limits_{l_0}^{l_p} dl
\int\limits_{l_0}^l dl'\,\hat q(l').
\label{320}
\eeq
Although this is a small correction, it is included in the calculations by making a proper shift of the variable $z_h$ in the fragmentation function.

\begin{figure}[htb]
\includegraphics[width=5.8 cm]{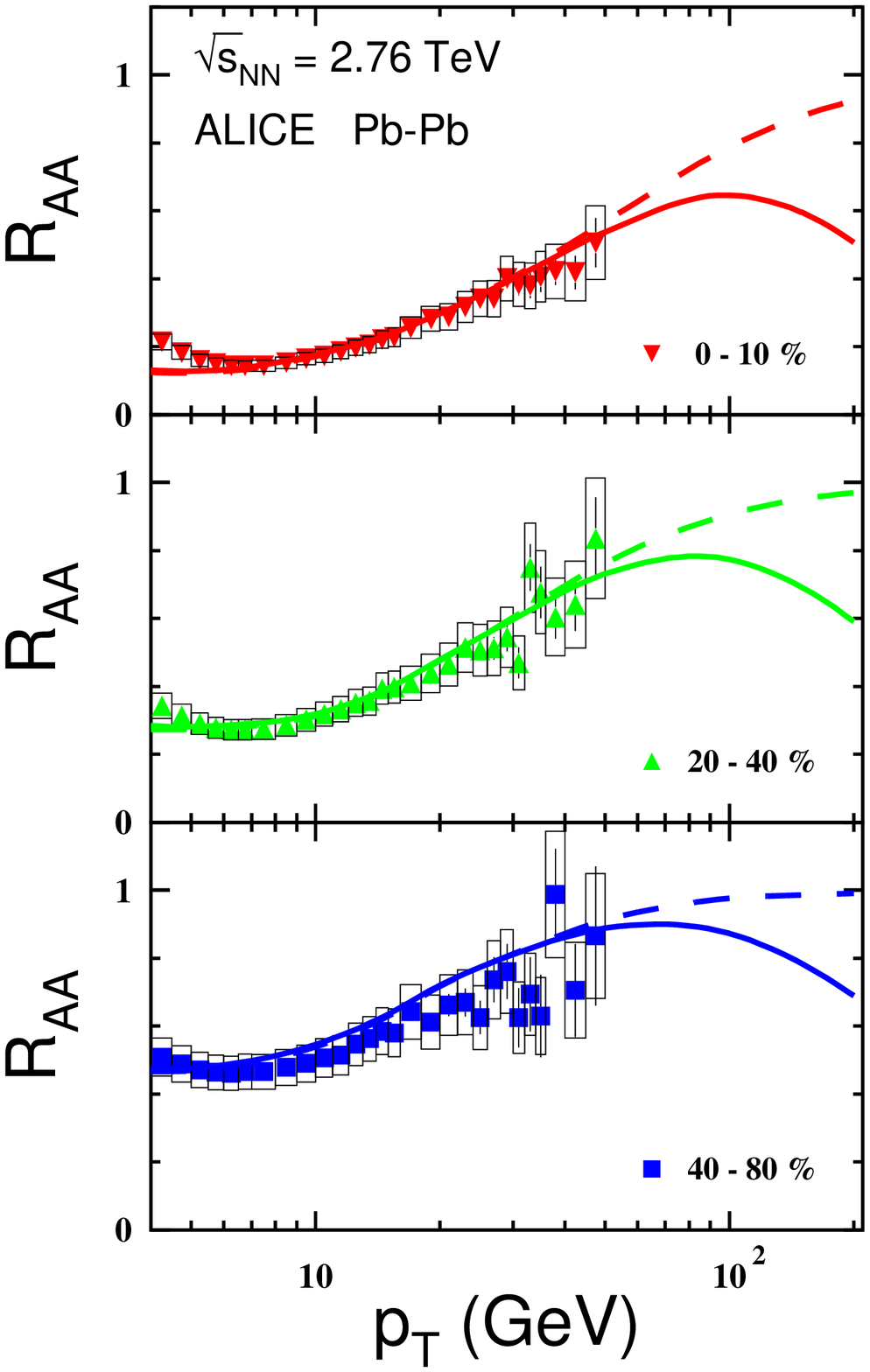}
\hspace*{-10mm}
\includegraphics[width=9cm]{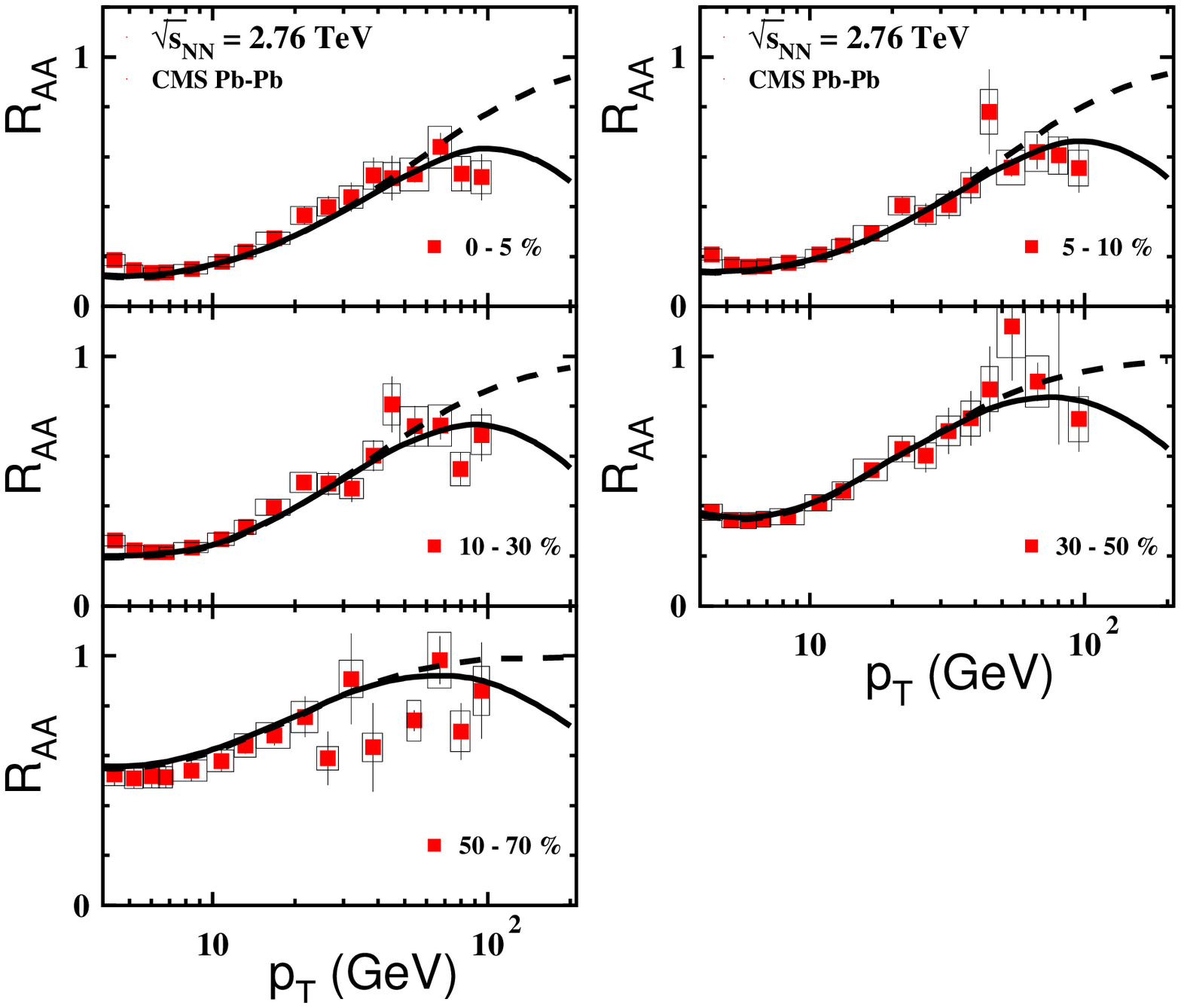}
\caption{ Centrality dependence of the suppression factor $R_{AA}(p_T,b)$ measured in the ALICE \cite{alice-new} ({\sl left}) and CMS \cite{cms-new1,cms-new2} ({\sl right}) experiments at $\sqrt{s}=2.76\TeV$. The intervals of centrality are indicated. 
The dashed lines are calculated with Eq.~(\ref{300}).
The solid curves also include the effects of initial state interactions \cite{isi,kn-review}, 
as is described in Sect.~\ref{isi}. 
\label{fig:raa-b-gf}}
\end{figure}

The results of calculations \cite{ct-eloss} for charge hadron suppression factor $R_{AA}(p_T)$ are plotted in Fig.~\ref{fig:raa-b-gf} in comparison with data from the ALICE \cite{alice-new} and CMS \cite{cms-new1,cms-new2} experiments at $\sqrt{s}=2.76\TeV$ and different  centralities indicated in the plot. 
The dashed lines are calculated with the path-integral expression, Eq.~(\ref{300}), calculated with the space- and time dependent transport coefficient Eq.~(\ref{280}), where the parameter  $\hat q_0=2\GeV^2\!/\!\fm$, which controls the normalization of $R_{AA}$, was adjusted to data. Notice that this parameter is independent of $p_T$ and centrality. It affects the normalization, but not the shape of the $p_T$ dependence.
The solid curves also include the effects of initial state interactions (ISI) and energy conservation in nuclear collisions \cite{isi,kn-review}, 
as is described below in Sect.~\ref{isi}.  

\begin{figure}[htb]
 \hspace*{-3mm}
\includegraphics[width=7.5 cm]{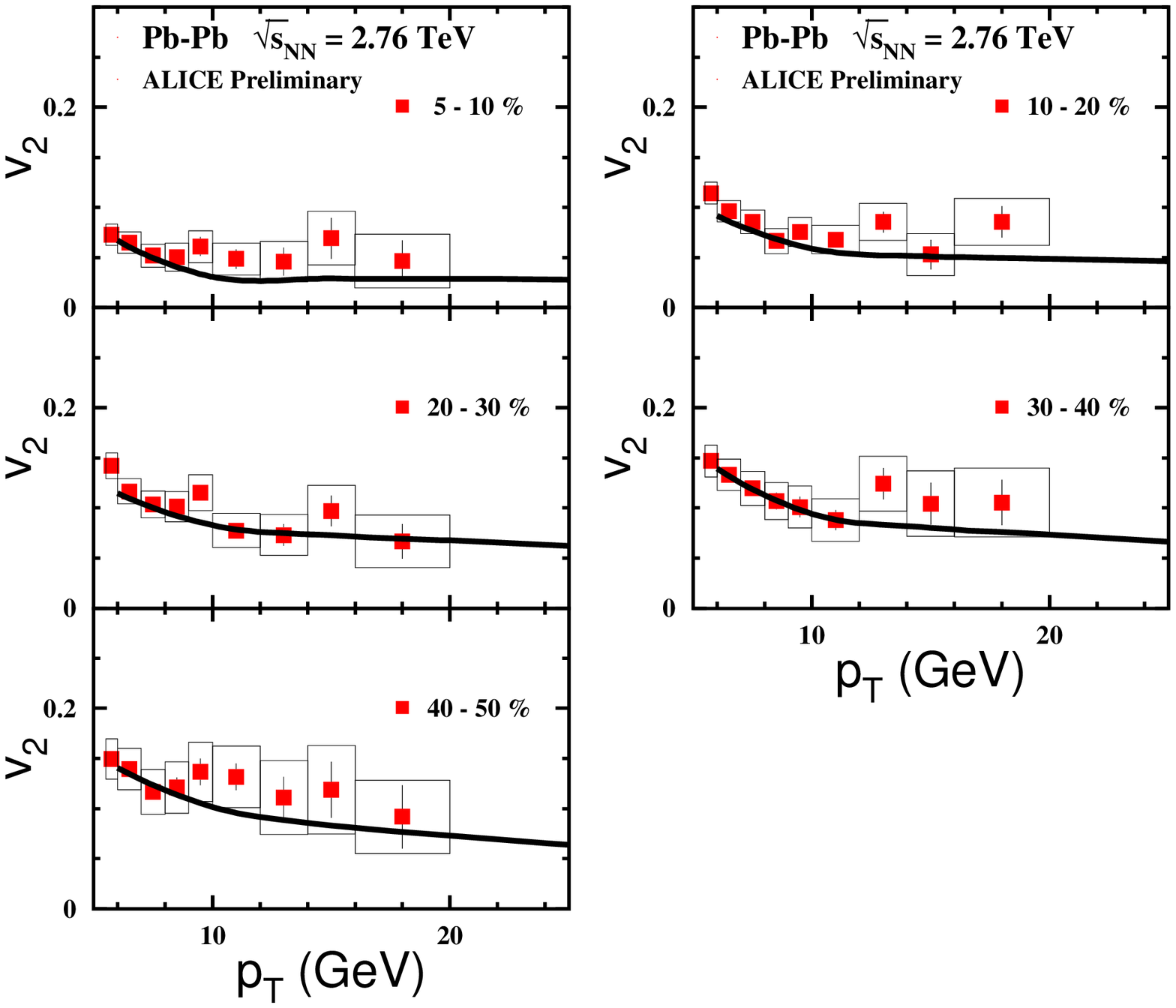}
\hspace*{-5mm}
\includegraphics[width=7.5 cm]{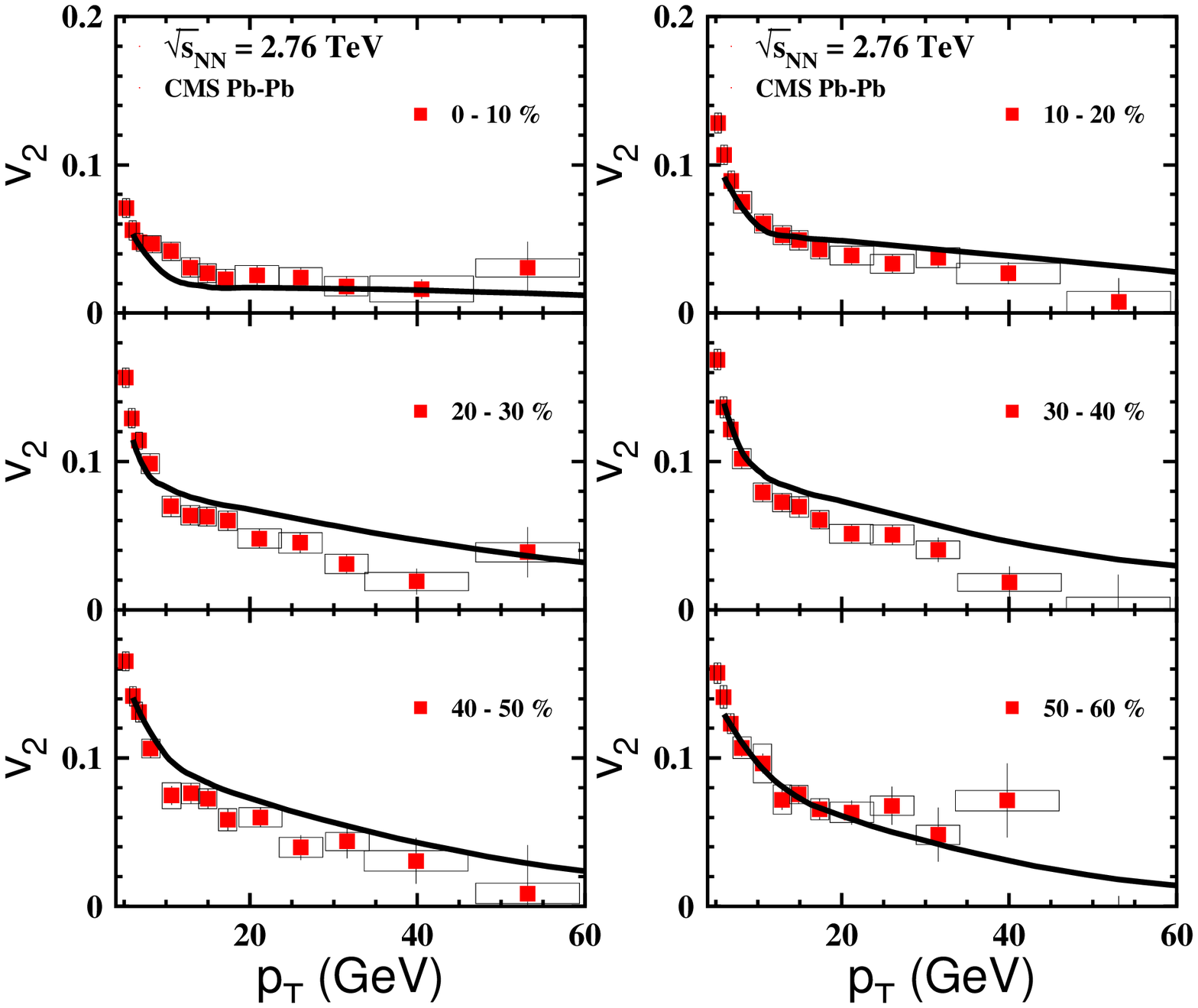}
\caption{ Data from the ALICE \cite{alice-phi-v2} ({\sl left}) and  CMS \cite{cms-v2} ({\sl right}) experiments for azimuthal anisotropy, $v_2(p_T)$ in lead-lead collisions at
$\sqrt{s}$ = 2.76 $\TeV$ and different
centralities. The curves present the results of calculation with Eq.~(\ref{340}).
\label{fig:v2-b-gf}}
\end{figure}

As far as hadrons propagated over a longer path are suppressed more, naturally the azimuthal angle distribution of the produced hadrons correlates with the geometry of the collisions and gains an asymmetry. Data for such an asymmetry, characterized by the parameter $v_2=\la \cos(2\phi)\ra$, provide an alternative sensitive way to test the model for suppression. The asymmetry can be calculated in a way similar
to Eq.~(\ref{300}),
\beq
v_2(p_T,b)=
\frac{\int d^2\tau \,T_A(\tau)T_B(\vec b-\vec\tau)\int\limits_0^{2\pi}d\phi
\cos(2\phi)\,
\left|\int\limits_0^\infty dr\,r\,\Psi_h(r) 
G_{\bar qq}(0,0;\infty, r)\right|^2}
{\int d^2\tau \,T_A(\tau)T_B(\vec b-\vec\tau)
\int\limits_0^{2\pi}d\phi 
\left|\int\limits_0^\infty dr\,r\,\Psi_h(r)\, 
G_{\bar qq}(0,0;\infty, r)\right|^2}.
\label{340}
\eeq
Here we neglected the initial dipole size $r_1\sim1/p_T\approx 0$ at $l_1=0$.
The results of calculations \cite{ct-eloss}  are compared in Fig.~\ref{fig:v2-b-gf}  with data from the ALICE \cite{alice-phi-v2} and CMS \cite{cms-v2} experiments at $\sqrt{s}=2.76\TeV$ and different  centralities. 
 The calculated asymmetry in the perturbative regime is rather small and is falling monotonically with $p_T$ in agreement with data.
 
Notice that an azimuthal asymmetry  appears for any mechanism of suppression.
One may wonder if a successful description of the cross section (Fig.~\ref{fig:raa-b-gf}) should automatically lead to a good agreement with $v_2(p_T)$, i.e. whether
data for $v_2$ suggest a complementary test of the model? The answer is positive, $v_2$ data provide a more stringent examination of the model. Indeed, while the suppression of
the cross section depends on the accumulated suppression along the hadron path in the medium, azimuthal asymmetry is sensitive to the path length distribution.

\section{\boldmath Large $x_T$, towards the kinematic bound}
\label{isi}

Multiple interactions of the projectile hadron and its debris propagating through the nucleus lead to 
a dissipation of energy. Important observation made in \cite{isi} (see also \cite{peigne}) is that the resulting loss of energy is proportional to the energy of the projectile hadron, therefore the related effects do not disappear at very high energies. An easy and intuitive way to understand this is the Fock state representation for the projectile hadron wave function.  Those sates are the fluctuations of this hadron
"frozen" by Lorentz time dilation.  The interaction with the target
modifies the weights of the Fock states, some interact stronger, some weaker. An example is the 
light-cone wave function of a transversely polarized photon \cite{krt2}. In vacuum it is overwhelmed 
by $\bar qq$ Fock states with vanishingly small separation (this is why the normalization of the 
wave function is ultraviolet divergent). However, those small size fluctuations have a vanishingly 
small interaction cross section, and the photoabsorption cross section turns out to be finite.

In each Fock component the hadron momentum is shared by the constituents, and the momentum 
distribution depends on their multiplicity: the more constituents the Fock state contains, the smaller is the 
mean fractional momentum per a constituent parton. Higher Fock components interact with a nuclear target stronger and gain larger weight factors compared to low Fock states. Thus, the $x$ distribution in the projectile hadron is softer on a nuclear than on a proton targets.

In the case of a hard reaction on a nucleus, such softening of the momentum 
distribution of a projectile parton can be viewed as an effective loss of energy of the leading parton due to initial state multiple interactions:  the mean energy of the leading parton on a nuclear target decreases compared the hard reaction on a proton target. Such a reduction of the fractional momentum of the 
leading parton is apparently independent of the initial 
hadron energy. Thus, the effective loss of energy is proportional to the initial energy.

Notice that this is different from energy loss by a single parton propagating 
through a medium and experiencing induced gluon radiation.
In this case the mean fractional energy carried the radiated gluons vanishes at large initial energies $E$ 
as $\Delta E/E\propto1/E$ \cite{feri,bh,bdmps}. 

Initial state energy loss is a minor effect at high energies and mid rapidities. However, it may essentially suppress the cross section upon approaching the kinematic bound, either $x_L=2p_L/\sqrt{s}\to1$
or $x_T=2p_T/\sqrt{s}\to1$. Correspondingly, the proper variable, which controls this effect is
$\xi=\sqrt{x_L^2+x_T^2}$. The magnitude of suppression was evaluated in \cite{isi,kn-review}.
It was found that each of multiple interactions (treated within the Glauber approximation) in the nucleus
supplies a suppression factor $U(\xi)\approx 1-\xi$. Summing up over the multiple ISI interactions
in $pA$ collision with impact parameter $b$
one arrives at a new parton distribution function in the projectile proton compared with $pp$ collisions,
 \beq
\hspace*{-0.40cm}
F^{(A)}_{i/p}(x,Q^2,b)=C\,F_{i/p}(x,Q^2)\,
\frac{
\left[e^{-\xi\sigma_{eff}T_A(b)}-
e^{-\sigma_{eff}T_A(b)}\right]}
{(1-\xi)\,\left[1-
e^{-\sigma_{eff}T_A(b)}\right]}.
\label{360}
 \eeq
Here $\sigma_{eff}$ is the effective hadronic cross section controlling multiple interactions. 
It is reduced by Gribov inelastic shadowing, which makes the nuclear medium more transparent. The effective cross section was evaluated  in \cite{isi,lrg} at about $\sigma_{eff}\approx 20\mb$.
 The normalization factor $C$ in Eq.~(\ref{360}) is fixed by the Gottfried
sum rule.
  
With the parton distribution functions Eq.~(\ref{360}) modified by ISI energy loss
one achieves a good parameter-free description of available data from the BRAHMAS \cite{brahms} and STAR \cite{star-forward} experiments at forward rapidities in $dA$ collisions large $x_L$ \cite{isi,kn-review}.

\begin{figure}[htb]
\parbox{\halftext}{
\vspace{3mm}
\centerline{\includegraphics[width=5.5 cm]{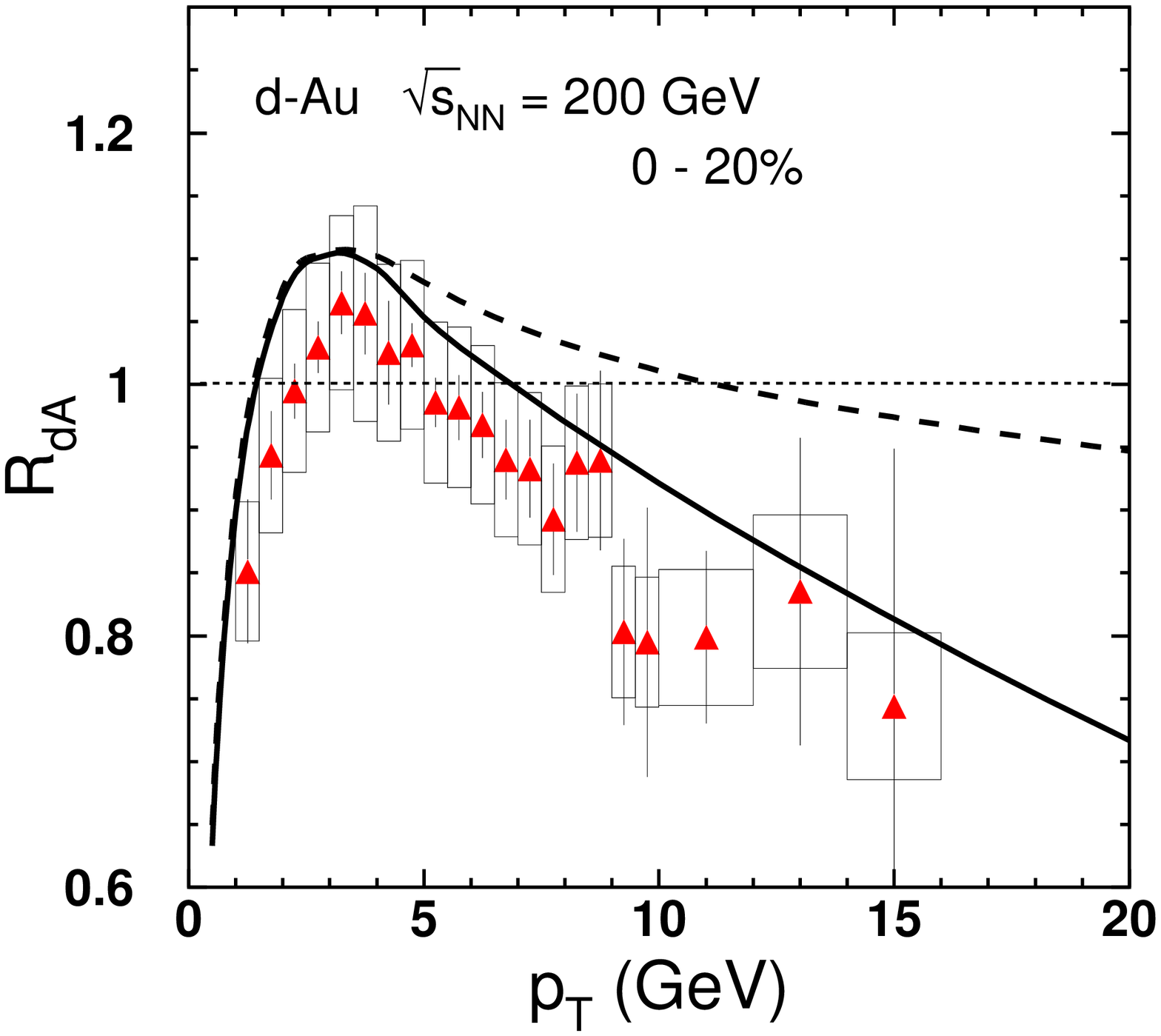}}
\caption{Nuclear attenuation factor $R_{dAu}(p_T)$ for  $\pi^0$ mesons produced in central ($0-20\%$) $d$-$Au$ collisions
at $\sqrt{s}=200\GeV$ and $\eta = 0$. The solid and dashed curves
show the  predictions calculated with and without the ISI corrections.  Isotopic effect is included. The data are from the PHENIX experiment \cite{cronin-phenix}.}
\label{fig:cronin}}
\hfill
\parbox{\halftext}{
\centerline{\includegraphics[width=5.5 cm]{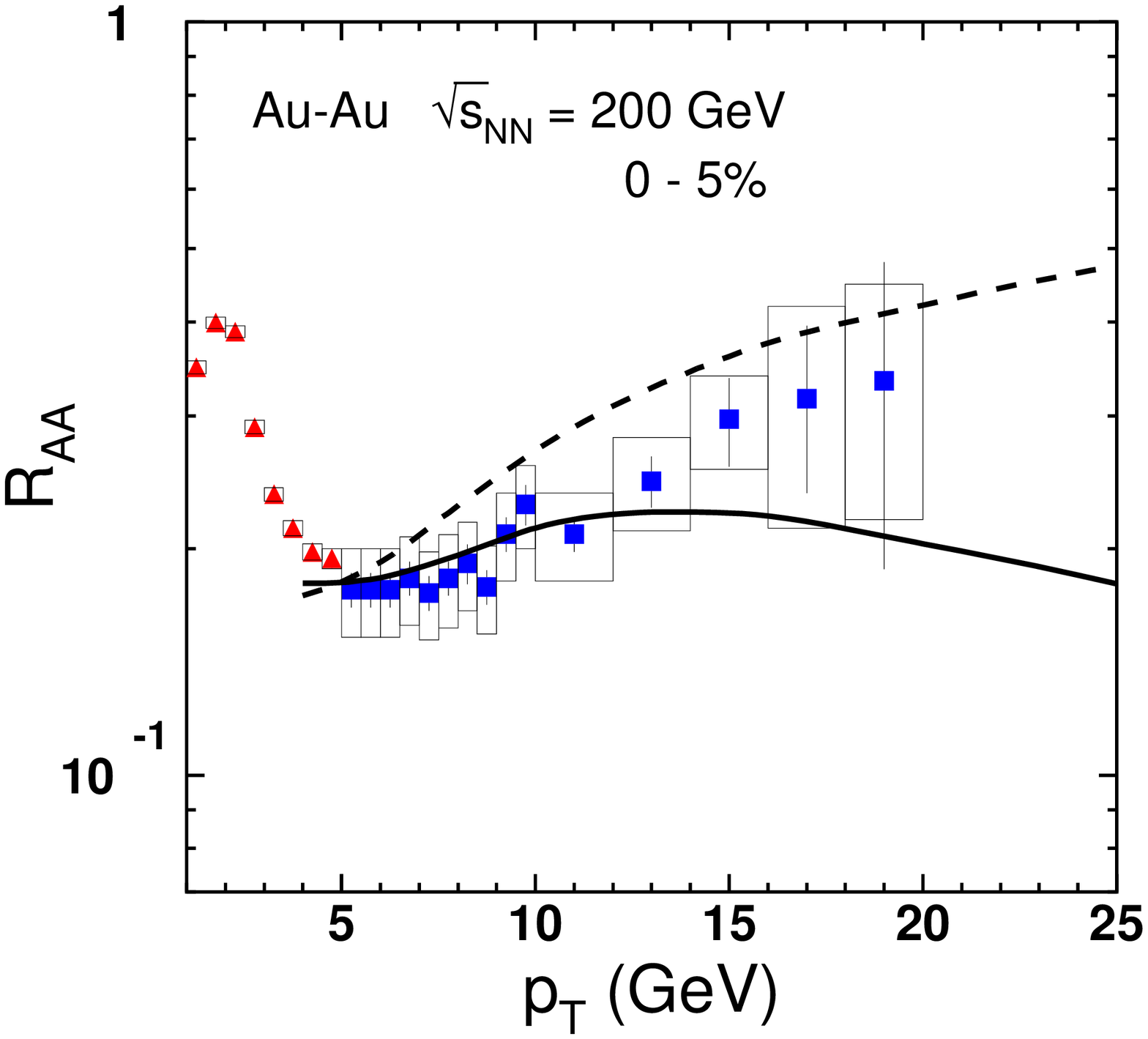}}
\caption{ Nuclear attenuation factor $R_{AA}(p_T)$ 
for neutral pions produced in central gold-gold collisions at
$\sqrt{s}$ = 200 $\GeV$.
The solid and dashed lines are calculated with or without
ISI corrections.
PHENIX data are from \cite{phenix-b} (triangles)
and \cite{phenix-0} (squares). } \label{fig:rhic-0-5}}
\end{figure}

The ISI energy-loss  should also be important at large $p_T$, in particular in the RHIC energy range, where $x_T$ in data reaches values of $0.2-0.3$. 
Notice that the real values of $x_T$, essential for energy conservation, are about twice larger,
$\tilde x_T=x_T/z_h$, so it reaches values of $0.5$ at RHIC, and about $0.4$
at LHC. The measured nuclear modification of pions produced at high $p_T$ in $dAu$ collisions at $\sqrt{s}=200\GeV$ \cite{cronin-phenix} indeed demonstrates a significant suppression, as one can see in Fig.~\ref{fig:cronin}. With the same modification factor Eq.~(\ref{360}), which was successful at forward rapidities, a good agreement is achieved at large $p_T$ as well, as is demonstrated in Fig.~\ref{fig:cronin}.

 \vspace*{-1cm}

The effects of ISI energy loss also affect the $p_T$ dependence of the nuclear suppression in heavy ion collisions. These effects are calculated similarly, using the modified 
parton distribution functions Eq.~(\ref{360}) for nucleons in both colliding nuclei.
The resulting additional suppression significantly reduces $R_{AA}(p_T)$ at the energy of RHIC.
This is demonstrated in Fig.~\ref{fig:rhic-0-5} on the example of central gold-gold collisions at $\sqrt{s}=200\GeV$. Since parameter $q_0$ is expected to vary with energy, it was readjusted and found to be $q_0(RHIC)=1.6\GeV^2\!/\!\fm$,  less than in collisions at the LHC\footnote{Notice that $q_0$ is also $A$-dependent.}.
Data on centrality dependence of $R_{AA}(p_T)$,  presented in Fig.~\ref{fig:rhic-0-5},
is also well explained.

 \begin{figure}[htb]
\centerline{
\includegraphics[width=5.5 cm]{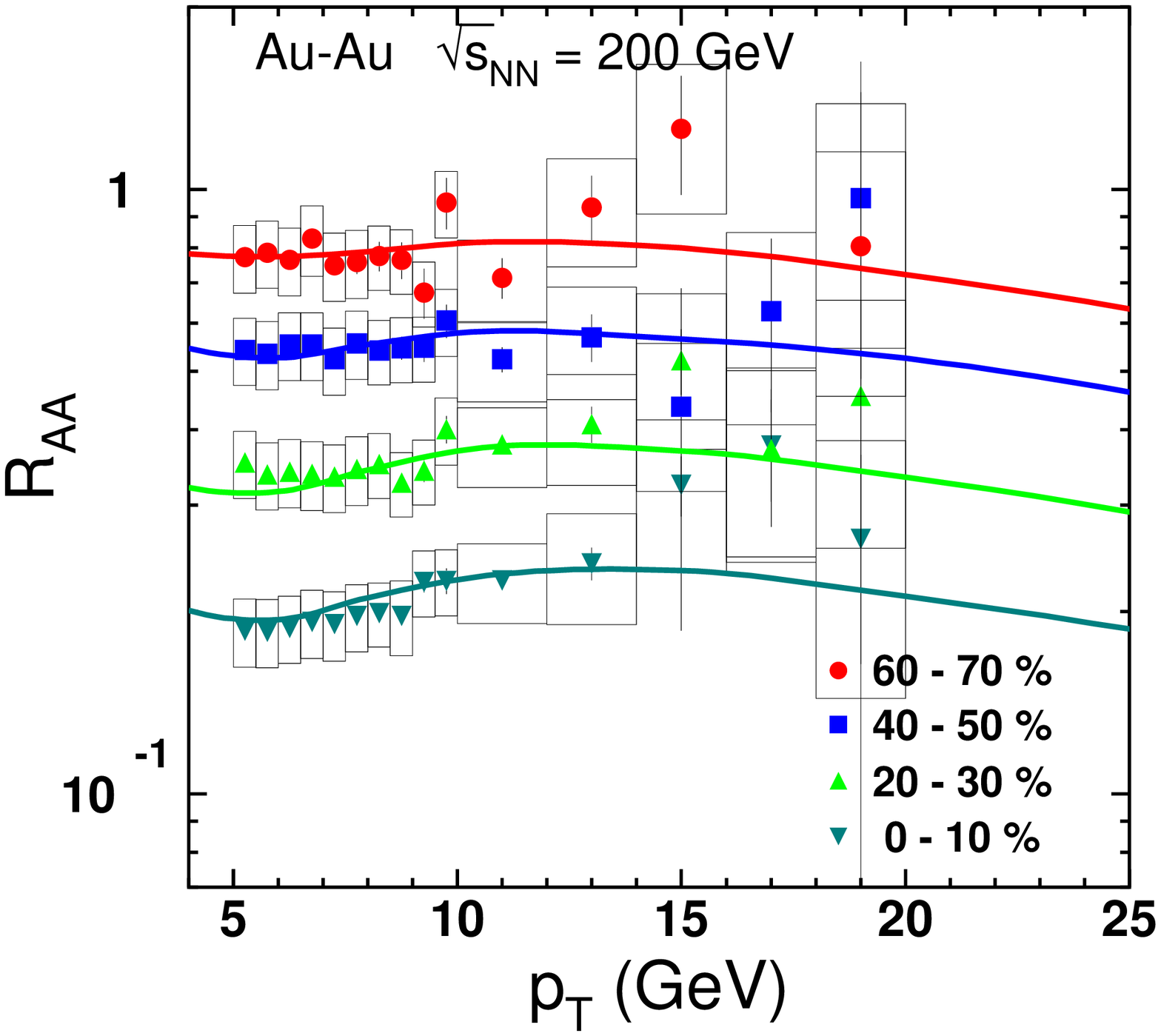}
\hspace*{1cm}
\includegraphics[width=5.5 cm]{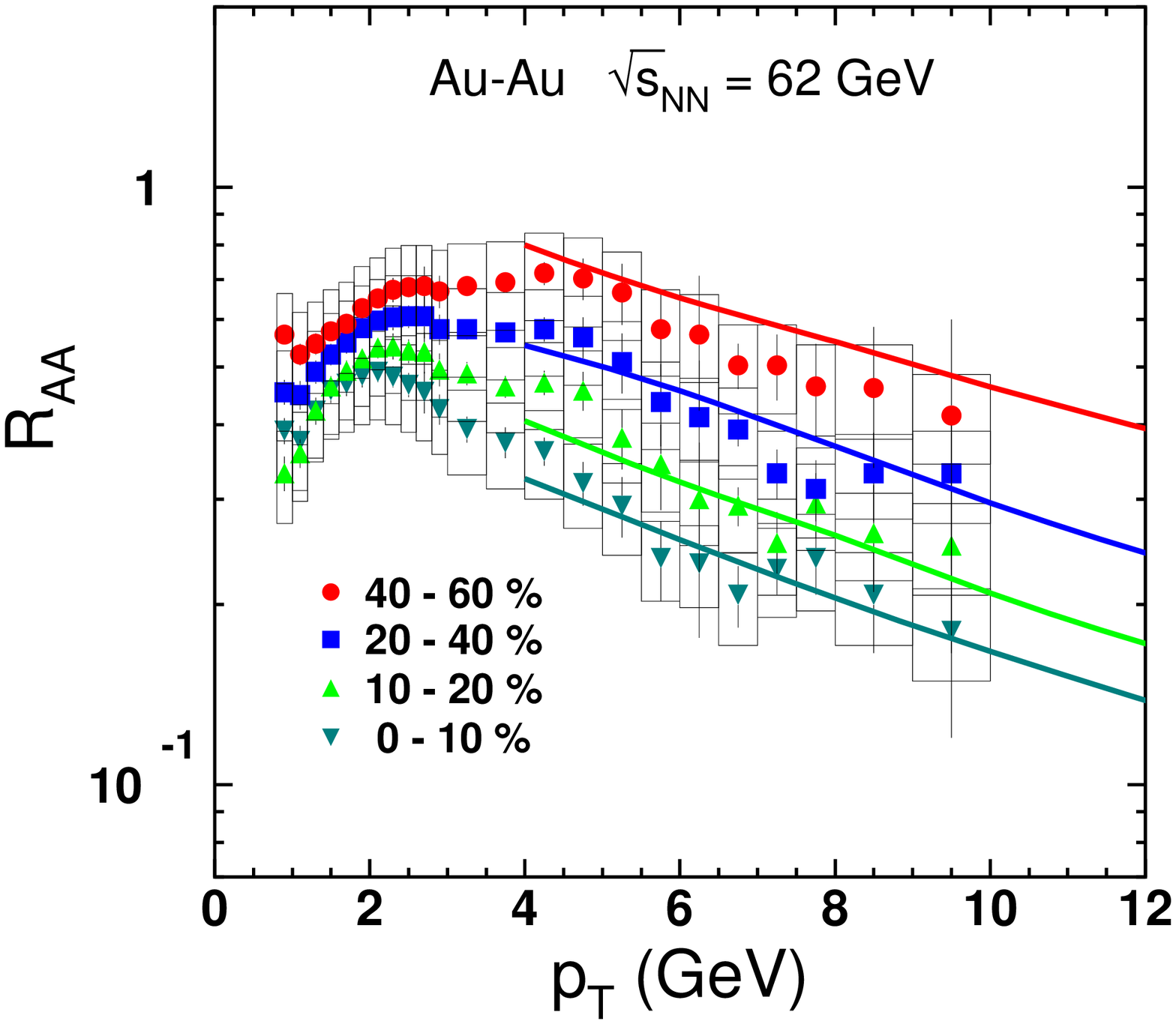}}
\caption{Centrality dependence of the suppression factor $R_{AA}(p_T,b)$ measured in the PHENIX experiment in gold-gold collisions at $\sqrt{s}=200\GeV$ \cite{phenix-b}({\sl left}) and
$\sqrt{s}=62\GeV$ \cite{phenix62-b} ({\sl right}). The intervals of centrality are indicated in the plot.
\label{fig:rhic62-b}}
\end{figure}

One can access larger values of $x_T$ either by increasing $p_T$ and a fixed energy, or by reducing the collision energy keeping unchanged the $p_T$ range. So the data \cite{phenix62-b} at lower energies $\sqrt{s}=62\GeV$ depicted in Fig.~\ref{fig:rhic62-b} play important role for study of the energy loss effects.
Indeed, the data demonstrate an unusual falling $p_T$ dependence of $R_{AA}(p_T)$ predicted with the suppression factor Eq.~(\ref{360}).

On the contrary, at much higher energies of LHC one would not expect any sizable effects of energy loss. Nevertheless, even at such high energies one can reach  $x_T$ sufficiently large for ISI energy loss effects to show up. The results depicted by solid curves in Fig.~\ref{fig:raa-b-gf}
include such energy loss corrections, which cause leveling off and even fall of $R_{AA}(p_T)$
at $p_T\gtrsim 100\GeV$.

\section{Summary}

Summarizing, for the process of inclusive high-$p_T$ hadron production we performed evaluation of the production length $l_p$ available for 
gluon radiation and energy loss, and found it to be rather short in many instances.
As a result, the main reason for the observed suppression in heavy ion collisions is not induced energy loss, but attenuation of early produced color dipoles propagating through a dense absorptive matter. Having no free parameters, except the medium density characterized by the transport coefficient $\hat q$, we reached a good agreement with data on
nuclear suppression and azimuthal elliptic flow in a wide range of energy, from the lowest energies of RHIC up to LHC,
and in a wide range of transverse momenta, from $p_T=5$-$7\GeV$ up to $100\GeV$.
The region of smaller $p_T$ apparently is dominated by hydrodynamic mechanisms of hadron production. The recent attempt \cite{hydro} to unify hydrodynamics with perturbative QCD calculations, presented above, was successful, the whole range of $p-T$ was well described with the same
medium temperature.

Our analysis led to quite reasonable values of the parameter characterizing the hot medium,
$q_0=1.2,\ 1.6$ and $2\GeV^2\!/\!\fm$ at $\sqrt{s}=62,\ 200$ and $2760\GeV$ respectively. This is close to the expected magnitude $q_0\sim 1\GeV^2\!/\!\fm$ \cite{bdmps}, as well as to the value extracted from data on $J/\Psi$ suppression \cite{psi,psi-bnl}. The latter is an alternative probe for the created hot matter, and
different probes obviously must result in the same properties of the probed medium.
The pure energy loss scenario did not pass this important test, it leads to a magnitude of the transport coefficient \cite{phenix-theor}, which
 is an order of magnitude larger than expected \cite{bdmps}.

Another important test of the energy loss scenario would be a direct measurement of the transport coefficient, i.e. broadening of a jet propagating through the hot medium.
On the contrary to this expectation no broadening was found in back-to-back photon-jet azimuthal correlation
\cite{cms-jets}.

Concluding, few words of precotion to avoid further confusions:

(i) One should clearly distinguish between vacuum and medium-induced radiative energy loss.
The former is much more intensive and is the main reason for shortness of $l_p$.

(ii) One should also discriminate between the terms "jet quenching" and "hadron quenching".
The latter is what was considered in this paper, a hadron detected inclusively with a high $p_T$
carries the main fraction $z_h$ of the accompanying jet. This is why intensive vacuum energy loss and energy conservation impose severe constraints on the value of $l_p$.
If, however, the whole jet has a large $p_T$, while none of hadrons in the jet are not forced to have large $z_h$, the hadronization length is subject to the usual Lorentz time dilation and is long.
 \\

{\bf Acknowledgments:}
This work was supported in part
by Fondecyt (Chile) grants 1130543, 1130549, 1100287, 
and by Conicyt-DFG grant No. RE 3513/1-1.
The work of J.N. was partially supported
by the grant 13-02841S of the Czech Science Foundation (GA\v{C}R),
by the Grant VZ M\v{S}MT 6840770039,
by the Slovak Research and Development Agency APVV-0050-11 and
by the Slovak Funding Agency, Grant
2/0092/10.

\end{document}